**Causally-interpretable meta-analysis: clearly-defined causal effects and two case studies**

Kollin W. Rott[1]

Gert Bronfort[2]

Haitao Chu[1]

Jared D. Huling[1]

Brent Leininger[2]

Mohammad Hassan Murad[3]

Zhen Wang[3]

James S. Hodges[1]

[1] Division of Biostatistics, University of Minnesota School of Public Health, Minneapolis MN 55455, USA

[2] Earl E. Bakken Center for Spirituality & Healing, University of Minnesota, Minneapolis MN 55455, USA

[3] Evidence-based Practice Center, Mayo Clinic, Rochester MN 55902, USA

**Abstract**

Meta-analysis is commonly used to combine results from multiple clinical trials, but traditional meta-analysis methods do not refer explicitly to a population of individuals to whom the results apply and it is not clear how to use their results to assess a treatment's effect for a population of interest. We describe recently-introduced causally-interpretable meta-analysis methods and apply their treatment effect estimators to two individual-participant data sets. These estimators transport estimated treatment effects from studies in the meta-analysis to a specified target population using individuals' potentially effect-modifying covariates. We consider different regression and weighting methods within this approach and compare the results to traditional aggregated-data meta-analysis methods. In our applications, certain versions of the causally-interpretable methods performed somewhat better than the traditional methods, but the latter generally did well. The causally-interpretable methods offer the most promise when covariates modify treatment effects and our results suggest that traditional methods work well when there is little effect heterogeneity. The causally-interpretable approach gives meta-analysis an appealing theoretical framework by relating an estimator directly to a specific population and lays a solid foundation for future developments.

## Introduction

A systematic review and meta-analysis is the preferred tool for evidence synthesis that supports decision-making.[1-3] The advantages of this tool include the ability to synthesize the whole body of evidence and that oftentimes it produces more precise estimates than those of individual studies and it provides an opportunity to reduce publication bias[4] and generalize the results to a broader population.[5] Treatment decisions are, however, inherently linked to the causal treatment effect in a specific population, which a standard meta-analysis may not be able to estimate. It is therefore of interest to make causally-interpretable generalizations from a group of studies to a new target population. Meta-analysis has been criticized for its limited ability to make such generalizations, whether as a pairwise meta-analysis (i.e.., comparing two treatments), network meta-analysis (i.e., comparing multiple treatments), aggregate data meta-analysis (pooling study level estimates) or individual participant data (IPD) meta-analysis (pooling participant level data across multiple studies).[6-7]

Causal inference is a field in statistics that has not historically been applied to meta-analysis.[8] The importance of causal inference is evident. A large part of practical application of statistics is at its foundation focused on understanding causal relationships. This idea is especially key in the health sciences, where it is paramount to understand the causal effect on patients of treatments or interventions. The distinction between association and causation is oft-discussed and causal inference provides a valid framework for understanding what is needed to estimate causal treatment effects. Causal inference about meta-analytic treatment effects requires the ability to transport conclusions from a group of studies to a new target population. Although traditional meta-analysis methods can provide a way to generalize causal inferences to a new population, it is not always feasible because outcome data are pooled across different populations with different characteristics using weights derived from the precision of the study-specific estimates. If the treatment effect is heterogeneous with respect to characteristics that are distributed differently in the included studies, estimated causal effects may not be transportable to a target population.[6-7,9-11] Moreover, the traditional methods have focused primarily on producing an effect estimate, and limited attention has been paid to the population to whom the effect estimate applies. Therefore, frameworks for causal inference in meta-analysis have been proposed to combine data from individual participants of the included studies using baseline covariate data from a random sample from the target population.[9-10] These methods, however, have rarely been applied.[12-14]

This paper implements and extends a recently-developed causally-interpretable meta-analysis method that relies on IPD.[9-10] This approach differs in a fundamental way from standard IPD meta-analyses and differs from previous approaches for causal inference that use aggregate data.[5-6,14-21] By using IPD to produce transportable estimates, the methods discussed in this paper offer a chance to obtain better, more precise and representative estimates than would be possible with only aggregated data. We propose multiple adjustments to the estimation process to stabilize the estimators in cases where causal assumptions are not fully met. We apply these methods to two previously-published data sets from distinct subject matter areas and compare them to traditional methods, with the aims of introducing the new methods to a larger audience, examining their strengths and weaknesses, and motivating improvements and further developments. The Methods section reviews traditional meta-analysis methods (including meta-regression) and the causally-interpretable methods that are the focus of the paper, introduces the data sets, and shows how we evaluate the methods' performance. The Results section compares the methods' performance on the two data sets using our evaluation metric and considers ways to address problems that these applications show. In the Discussion section, we focus on the implications of the results and how to move forward with causally-interpretable meta-analysis.

## Methods

### Traditional meta-analysis methods

Traditional meta-analysis models include common-effect, fixed-effects, and random-effects meta-analysis.[22] Each model's weighting method and target parameter depend on the assumptions made about the effect-size parameters in individual studies. The common-effect approach assumes all studies share the same effect. This assumption is tenuous in nearly all cases given differences between clinical trials in protocols and populations. The common-effect approach is also sometimes called the fixed-effect approach, using the singular noun, not to be confused with the fixed-effects approach, using the plural noun, which assumes the study-level effect sizes are fixed but not necessarily the same. To provide a combined estimate of treatment effect, both of these approaches generally weight the study-level effect-size estimates using the inverse variance of the estimates, but they estimate different target parameters with different interpretations. The target parameter under the common-effect approach is "the effect in any combination of the individual study populations", while the target parameter under the fixed-effects approach applies to the population "formed by amalgamating the study populations at hand".[23] The third popular approach to pairwise meta-analysis, the random-effects approach, assumes the study-level effect sizes can differ and are drawn from a distribution (often but not necessarily normal). This approach explicitly models and quantifies between-study heterogeneity[24-26] and uses it in estimating the target parameter, the mean of the distribution of effect sizes, and assigning it a measure of uncertainty.

Although incorporating heterogeneity is attractive, the population to which this target parameter applies is difficult to understand and therefore has no useful clinical interpretation.[27] In fact, in all three cases, the estimated effect does not apply to any particular well-specified population. Arguably the clearest interpretation, that of the fixed-effects approach, simply defines the target population as an amalgamation of the included study populations. One can argue that the absence of a target population impairs the practical utility of the analysis. The goal in estimating a treatment effect is nearly always to determine how the treatment will work in a particular group of people who could receive the treatment in the future. Rather than estimating the treatment effect in a particular population, however, traditional meta-analysis estimators estimate the effect in a population that is at best nebulous and even potentially unknown.

As IPD have become more available in recent years,[28] interest has increased in extending the above methods to IPD. Briefly, methods using IPD can incorporate regression on characteristics in a more integral and meaningful way than methods using aggregate data.[15-17] These methods allow meta-analytic estimates of treatment effects to depend to some extent on individual characteristics. Much like methods using aggregate data, however, they do not refer to a specific target population.

Meta-regression, in which aggregated study-level covariates are included in the meta-analysis, may be used to make inferences about a target population while accounting for potential effect moderators[29]. Making predictions for a target population in the meta-regression context is uncommon but has been done recently[30]. There has been significant research to increase the practical utility of meta-regression, for example implementing machine learning approaches or using cross-validation for model evaluation[31,32]. Other research suggests an information-theoretic approach to improve model selection in meta-analysis models, which is not a focus of this methodological paper but will be of interest to those wishing to use such methods in practice[33]. Robust variance estimation, in which both between-study and within-study variance are properly accounted for, has also been proposed for meta-regression models[34]. In this paper, we consider for comparison a common and simple approach, a linear meta-regression model with study-aggregate covariate values. Meta-regression has the advantage of conceptual simplicity and broad availability of aggregate data. However, aggregate data cannot produce inferences on par with individual participant data when used correctly, as IPD allows for a better understanding of within-study variation and avoids the ecological fallacy inherent in meta-regression on aggregate data. We note that meta-regression focuses on a conditional treatment effect, i.e. conditional on study or participant characteristics, whereas the approach to meta-analysis based on causal inference focuses on a marginal treatment effect, which may be of more interest.

**Causal inference methods**

Hasegawa et al[7] introduce a target population into the traditional fixed-effects method by choosing study weights so that the weighted averages of the covariate means from the meta-analysis studies closely match the covariate means in the target population. Other approaches are based explicitly on ideas and methods from causal inference; we focus on the methods of Dahabreh et al[9-10,14], who proposed estimators to transport results from randomized trials in a meta-analysis to a clearly specified target population. This work builds on extensive work in causal inference for generalizability and transportability of causal effect estimates, as reviewed in Colnet, et al and Degtiar and Rose.[12-13] Related though not identical approaches are described by Kabali & Ghazipura[35] and Sobel, Madigan, and Wang[36].

The estimators proposed by Dahabreh et al use two distinct approaches common in causal inference. The cited papers give assumptions that rationalize the estimators and discuss these assumptions in detail; we briefly mention two of them here. First, outcomes are assumed to be exchangeable across studies conditional on the individuals' covariate values, although versions of the estimators exist that allow this assumption to be relaxed somewhat. Second, the probability of being in a meta-analysis study (as opposed to the target population), conditional on the individuals' covariate values, is assumed to be the same for individuals from all studies. These assumptions mean that the studies can be pooled together, ignoring potential between-study heterogeneity of treatment effects not captured by conditioning on covariates. Moreover, these assumptions imply that there are no omitted between-study confounds that might introduce omitted variable bias into the estimated treatment effect.

These assumptions are restrictive in their current form, which is a weakness of this nascent approach for meta-analysis. In the original paper by Dahabreh et al[9], the authors briefly outline an approach for sensitivity analysis of violations of these assumptions. In other work they gave a more detailed approach for the case of transporting a single trial to a target population[37]. These sensitivity analysis approaches are largely the focus of future work but serve as a warning about the potential effects of confounding in meta-analysis when not treated with care. Nonetheless, the causal framework on which this approach is built provides a principled foundation for further developments that can account for additional unexplained heterogeneity of effects, possibly using the conventions and setup of VanderWeele and Hernán.[38-40] That setup allows a mathematically rigorous approach in which heterogeneity between studies is built into the counterfactuals of the individuals in the studies; ongoing work focuses on extending the approach of Dahabreh et al to accommodate this variety of heterogeneity.

*Outcome model approach*[9-10]

The first approach, the *outcome model approach*, estimates the outcome for an individual in the target population, conditional on their covariate values, using the individuals in the meta-analysis studies. The approach then averages that conditional estimate against the distribution of covariates values in a sample from the target population to estimate the average treatment effect in the target population. Specifically, this estimator can be written as

$$\hat{\phi}(a, a') = \left\{\sum_{i=1}^{n} I(S_i = 0)\right\}^{-1} \sum_{i=1}^{n} I(S_i = 0)\{\hat{g}_a(X_i) - \hat{g}_{a'}(X_i)\}$$

where $a$ and $a'$ denote the two treatments being compared, $I([\text{condition}])$ is the indicator function taking the value 1 if [condition] is true and 0 otherwise, $S_i \in S$ is the study from which each of the $n$ individuals comes (the target population is labeled study 0), $X_i$ is individual $i$'s covariate values and $\hat{g}_a(X_i)$ is an estimator for $E[Y_i \mid X_i, S_i \in S, A_i = a]$, where $Y_i$ is the outcome for individual $i$ and $A_i$ is individual $i$'s treatment assignment. The function $\hat{g}_a(X_i)$ can be estimated in any of a variety of ways including fully-parametric, semi-parametric, or non-parametric approaches. We note this approach's similarity to two methods outside the causal-inference literature, simulated treatment comparison[41] and multilevel network meta-regression for population-adjusted treatment comparisons[42], while also noting that these methods lack an explicit causal-inference formulation.

(References 9 and 10 use somewhat different assumptions to identify an estimand, resulting in somewhat different versions of this estimator. Reference 9's version directly estimates the difference between the two treatments; reference 10's version, shown here, estimates the average outcome in each treatment group, and estimates the treatment effect as the difference between the two groups' average outcomes. Reference 10 discusses why that version is now preferred by the original authors; it also seems desirable to us to estimate mean outcomes in each treatment group rather than solely the treatment effect.)

*Inverse probability weighting (IPW) approach*

The second approach, *inverse probability weighting*, weights individuals in the meta-analysis studies based on how similar their covariates are to the covariate distribution in the target population. Individuals resembling those in the target population most closely are weighted the most. The estimator uses these weights and the individuals' outcome values to compute a weighted average estimate of the average treatment effect in the target population. The estimator can be written as

$$\hat{\phi}_w(a, a') = \left\{\sum_{i=1}^{n} I(S_i = 0)\right\}^{-1} \sum_{i=1}^{n} \left(\frac{I(A_i = a)}{\hat{e}_a(X_i)} - \frac{I(A_i = a')}{\hat{e}_{a'}(X_i)}\right) I(S_i \in S) \frac{\hat{p}(X_i)}{1 - \hat{p}(X_i)} Y_i$$

where $\hat{p}(X_i)$ estimates $\Pr[S = 0|X_i]$, the probability an individual is in the target population conditional on their covariate values, and $\hat{e}_a(X_i)$ estimates $\Pr[A_i = a \mid X_i, I(S_i \in S) = 1]$, the probability an individual receives the specified treatment conditional on their covariate values and being in a meta-analysis study. The inverse probability weights reweight the distribution of covariates in the collection of studies to approximate the distribution of covariates in the target population. We note this method's similarity to matching-adjusted indirect comparisons[43] (MAIC), but also that MAIC lacks an explicit causal-inference formulation and does not extend readily beyond comparing two treatments.

(Similar to the outcome model approach, references 9 and 10 use different assumptions to identify somewhat different estimands. In this case, the practical difference is how to estimate $\hat{e}_a(X_i)$; reference 9 estimates the probability of treatment assignment conditional on the meta-analysis study from which the individual comes, whereas reference 10, used here, estimates the probability conditional only on being in a meta-analysis study. The different versions respectively estimate the difference between the two treatment effects either directly or as a difference of the two groups' average outcomes, much like in the outcome model approach.)

*Computing estimates using the two methods*

The code provided in the original paper[9] used M-estimation as implemented in the R package "geex"[44] to simultaneously estimate, in the IPW method, $\hat{p}(X_i)$ and the treatment effect in the target population, and then to compute a sandwich variance estimate for the estimated treatment effect. That paper also suggested estimating the variance using the bootstrap, which we do in this paper. Specifically, the code from the original paper used logistic regression in the context of M-estimation to estimate $\hat{p}(X_i)$. In cases where the target population and the meta-analysis study populations have limited overlap in covariates, this approach can fail or be unstable in a bootstrap because the logistic regression gives coefficient estimates that are infinite or nearly so. Cases of limited overlap potentially violate the positivity assumption, a condition required to identify the average causal treatment effect. It could be argued, however, that in practice these are the cases where causally-interpretable methods are most useful. If the target population and all study populations are very similar, then the traditional aggregated meta-analysis estimators described earlier could provide a reasonable estimate of the causal effect in the target population. Causal transportability could provide the largest benefit precisely when the target population and at least some of the study populations differ substantively.

To avoid the computational problems and parametric limitations of logistic regression, we estimate $\hat{p}(X_i)$ using non-parametric conditional density estimation as implemented in the R package "np".[45] This code allows manual or automatic bandwidth selection to smooth the estimated density, and therefore does not fail in the way logistic regression does. As shown below in the Results section, this approach stabilized estimates so that they are not orders of magnitude outside the observed data.

Nonparametric density estimation cannot fix all overlap problems. In the chiropractic data set described below, three studies enrolled individuals with back pain and three with neck pain; therefore, when estimating $\hat{p}(X_i)$ and treating one of the six studies as the sample from the target population, the conditional density estimation approach gave essentially zero weight to anyone from the three studies with the other type of pain, meaning that the model effectively used individuals from only the two studies with the same pain location.

Weighting methods for causal inference can often be improved by normalizing the weights within each treatment group to have mean one. This normalization, when used in an estimate, is called a Hajek estimator[46] and can improve estimate variability. When the weights are positive, this normalization forces estimates to be sample bounded, guaranteeing that the resulting estimates interpolate the data, which the outcome model cannot guarantee. The adjustment can be useful in a case where the sum of the weights differs rather substantially from the actual sample size in the target population, although such a substantial difference is also an indication that some of the method's assumptions were violated (Issa Dahabreh, personal communication).

*Doubly-robust approach*

More recent work[10] has added to these models by changing some of the causal assumptions so that the potential outcome means can be identified and estimated consistently using a so-called doubly-robust approach that combines the two methods discussed above. The estimator for the potential outcome mean under treatment $a$ can be written as

$$\hat{\psi}(a) = \left\{\sum_{i=1}^{n} I(S_i = 0)\right\}^{-1} \sum_{i=1}^{n} \left\{ I(S_i \in S, A_i = a) \frac{\hat{p}(X_i)}{\hat{e}_a(X_i)(1 - \hat{p}(X_i))} \{Y_i - \hat{g}_a(X_i)\} + I(S_i = 0)\hat{g}_a(X_i)\right\}.$$

Then the estimated average treatment effect is the difference between the estimated potential outcome means under the two treatments, $\hat{\psi}(a) - \hat{\psi}(a')$. As a doubly-robust method, this approach is consistent as long as at least one of the outcome or inverse probability weighting models is correctly specified.

It is important to recognize that these transportability estimators improve upon traditional meta-analysis estimators only if the covariates modify the treatment effect. The standard aggregated approaches fail to produce causally-interpretable estimates "whenever treatment effects are heterogeneous over variables that vary in distribution across trials".[9] This result is intuitive and can be derived mathematically. Therefore, to use these methods, an important problem is choosing covariates on which to condition, i.e. to include in the model. Improper variable selection can lead to unexpected results, such as unstable weights in the inverse probability weighting method. For example, when analyzing the chiropractic data set described below, we initially included six covariates that were deemed of substantive interest. Using all six covariates, however, led to near-complete separation between the target population and the meta-analysis studies even using the conditional non-parametric density approach, with $\hat{p}(X_i)$ either 0 or 1 for nearly every individual. This leads to extremely large and unstable weights, which in turn leads to an unstable estimate of the average treatment effect. To avoid using too many covariates (which could introduce overlap problems), we did preliminary analyses of the meta-analysis data sets, using linear models, to find covariates that had significant interactions with the treatment, and included only those covariates in the models. This reduced dimensionality and mitigated the overlap problem while still using covariates that showed effect modification. The Supplement describes these preliminary analyses. We make no claim that our covariate-selection method is optimal or even very good; choosing such methods is an open research question, as seen in a previous citation about model selection in meta-analysis[33]. We recommend thinking carefully about how to select effect-modifying covariates in an applied problem.

**Data sets used in this paper**

We used two individual participant data (IPD) datasets, which we now describe.

*Real-time continuous glucose monitoring*

The first data set was used in a meta-analysis of the effect of real-time continuous glucose monitoring in individuals with type 1 diabetes.[47] We were able to include 8 studies having a combined sample size of 1371 with individual study sample sizes ranging from 111 to 433. The continuous outcome was HbA1c at follow-up, with three covariates: HbA1c at baseline, age, and sex (coded as binary). 7 of the studies enrolled mostly adults, with mean ages in the 20s and 30s, while the other study enrolled only children, with mean age 7. The weak age overlap between the latter study and the other studies gives an example case of the overlap problem discussed earlier.

*Chiropractic treatment*

The unpublished second IPD data set includes studies of the effect of various chiropractic procedures or exercise on back, leg, or neck pain. We were able to include 6 studies having a combined sample size of 984, with individual study sample sizes ranging from 143 to 190. The outcome was pain level after 12 weeks of follow-up, measured on a 0-10 integer visual analogue scale. The two covariates we chose, as described in the Supplement, were baseline pain (averaged across two baseline visits and thus on a 0-10 half-point scale) and pain location, a categorical variable.

These studies included more than two distinct treatments, which would ordinarily require a network meta-analysis. However, some of the groups are combinations of multiple treatments, so for the present purpose we labelled the groups so we could combine groups that shared certain treatments while ignoring the other parts of treatment. This relabeling achieves two goals: it provides a meta-analysis with a reasonable number of studies and it creates known study-level heterogeneity within treatment groups, although we do not know whether that heterogeneity extends to the treatment effects. Another problem is that each study was conducted entirely on populations with a single pain location (back, leg, or neck). We excluded a seventh study, the only study that included only people with leg pain, because of a complete lack of overlap with all other studies, which included those with either back or neck pain. Of the six remaining studies, three included persons with back pain and three with neck pain, with almost equal sample sizes.

**Evaluation metric and computing**

We assessed the methods by treating each MA study in turn as a sample from the target population and transporting the other studies to that "target", in a manner similar to leave-one-out cross-validation. Because we have the actual outcomes for individuals in the target study, we can calculate that study's estimated average treatment effect as an unbiased estimate of the true average treatment effect in that study population. We can then compare the transported estimates with this benchmark to measure the performance of each estimator. A natural approach to measure the performance is to take the absolute difference between the transported estimate and the observed average treatment effect for each target, and average that result across the targets. We did this, adjusting for sample size by dividing each study's absolute difference by the contribution of the group sample sizes to the standard error of the treatment effect, assuming a pooled estimate of within-group standard deviation:

$$SAD = \sum_{m=1}^{M} |\hat{\phi}_m - \phi_{obs,m}| \left(\frac{1}{n_{A,m}} + \frac{1}{n_{A',m}}\right)^{-\frac{1}{2}}$$

where the $n$ terms are the sample sizes in the labeled groups and studies.

This average standardized absolute difference metric is used to compare performance of the methods.

The Supplement gives example R code for the analyses. As explained, the base R functions "lm" and "glm" and/or the package "geex" are used for the estimators as computed in the original paper, while the R package "np" is used to replace those models with non-parametric conditional density estimates when appropriate.[44] The R package "meta" is used to compute the traditional fixed- and random-effects aggregated meta-analysis estimates, along with the meta-regression estimates using the function "metareg".[48]

## Results

*Real-time continuous glucose monitoring*

We applied the estimators proposed by Dahabreh et al to the continuous glucose monitoring dataset. We also calculated the traditional fixed-effect and random-effect estimates using the aggregated data. Table 1 shows the estimates for each of these meta-analysis methods, treating each study population in turn as the target population, along with the actual estimated treatment effect in the target study. Figure 1 shows the estimates and associated 95% confidence intervals. The outcome model (OM) approach proposed by Dahabreh et al[10] gave the smallest average standardized absolute difference from the observed treatment effect. The fixed-effect and random-effect aggregated meta-analysis methods gave nearly-identical average standardized absolute differences, and performed about 23% worse than the OM approach according to that measure. The aggregate meta-regression model performed worse than these methods. The non-Hajek versions of the inverse probability weighting (IPW) method performed much worse than the OM and traditional MA estimators. In particular, the version of IPW that uses logistic regression to predict whether an individual is in the target population sometimes estimated treatment effects more than an order of magnitude from the treatment effects observed in the target study. The extremely poor performance was due to logistic regression estimating fitted probabilities very close to zero or one, leading to unstable weights. Nonparametric conditional density estimation gave more stable individual weights, and IPW using this approach showed much better performance than using logistic regression, although the OM and traditional MA estimates still performed much better. The Hajek-normalized weighting estimators performed better than their unnormalized counterparts, preventing the extremely poor performance that occurred for the latter. In fact, the Hajek-normalized version of IPW using nonparametric conditional density estimation performed better than all methods except the OM approach in terms of standardized absolute difference.

We note that in this data set, there is not much evidence that the treatment effect depends on the covariates. (Our analysis is summarized in the Supplement.) Because the methods proposed by Dahabreh et al are intended to leverage effect modification, it is perhaps unsurprising that the traditional MA approaches performed nearly as well as the best implementations of the causally-interpretable approaches for a data set where effect modification appears modest at best.

To determine why the traditional meta-analysis estimates outperformed the unnormalized inverse probability weighting estimates, we rewrote the IPW method in the form of aggregated meta-analysis, with each study having an implied weight and a treatment-effect estimate. By construction, the weights in traditional aggregated MA (including fixed-effect and random-effect MA) must add to 1. When rewriting the IPW method in this form, the implied weights do not necessarily add to 1. If the sum of the implied weights differs substantially from 1, this indicates that covariate overlap between the target population and meta-analysis studies is problematic. For example, study 9 consisted entirely of children and had only partial age overlap with the other studies. When rewriting the IPW method in the form of an aggregated MA with study 9 as the target, the implied study weights summed to 0.56. Although this example is extreme, this rewriting exercise is useful to assess whether the studies have adequate covariate overlap, as assumed by the Dahabreh methods. The Hajek normalization fixes this problem, and rewriting the IPW method in this form can serve as a useful diagnostic for situations in which Hajek normalization is likely to help (Issa Dahabreh, personal communication).

We further investigated the performance of the various MA methods by visual inspection of the weights assigned to individuals. This inspection allowed us to notice that even for target studies for which covariate overlap was not a problem (e.g., Study 4), a long upper tail in the distribution of the covariate HbA1c led to a large portion of the weight being assigned to a relatively small number of individuals in that upper tail. We tried to mitigate the long tail by log-transforming HbA1c. Although this adjustment did make the variable somewhat more symmetric, it did not substantively change the results. The lesson from this problem is that even when reasonable overlap is present in the traditional sense, individual weights can be quite unstable due to problems specific to an individual data set that are not immediately obvious.

*Chiropractic treatment*

We also applied the traditional MA estimators and those proposed by Dahabreh et al to the chiropractic data set. Table 2 shows the estimates for each of these meta-analysis methods, treating each study population in turn as the target population, along with the actual estimated treatment effect in the target study. Figure 2 shows the estimates and associated 95% confidence intervals. The relative performance of the estimators differs for this data set compared to the glucose monitoring data set. The traditional MA estimators (that is, the fixed-effects and random-effects estimators) performed best in terms of average standardized absolute error, at least 15% better according to this measure, and gave nearly-identical estimates. Meta-regression again performed demonstrably worse than these approaches, perhaps suggesting an ecological bias. Among the causally-interpretable methods, the IPW approach using non-parametric conditional density estimation performed slightly better than its IPW counterpart with logistic regression, and the OM approach performed the worst. Again, we are presented with the question of why the traditional MA estimators seem to do so well without using covariate information of any kind and without focusing on a specific target population. Interestingly, when rewriting the IPW method in the form of an aggregated MA as discussed in the previous section, the sum of the implied weights for each target is very close to 1, suggesting that the Hajek normalization would not be that useful. In fact, the Hajek-normalized versions of the IPW methods changed results negligibly in this data set, making performance slightly worse.

**Discussion**

Perhaps the most striking result from applying the causally-interpretable estimators to two real individual participant data sets is that they did not outperform the traditional aggregate-data meta-analysis estimators by as much as might have been expected. Although this may seem discouraging, it can instead be seen as validating the traditional meta-analysis estimators within a causal framework in the case where included covariates modify the treatment effect only modestly at most. When there are no effect modifiers in the target patient population, the aggregate-data meta-analysis estimators estimate the average treatment effect in the target population even though effects are not explicitly transported. Even in cases where effect modification appeared to be present, albeit modest, the traditional estimators gave reasonable estimates compared to the causally-interpretable IPD estimators.

Another key result from this application is that different methods to compute weights can give substantially different results, especially when positivity assumptions may be violated. Unstabilized weights computed using logistic regression led to highly variable, poor estimates. Unstabilized weights using nonparametric conditional density estimation improved on the logistic regression approach but still gave highly variable and poor estimates. Stabilizing weights through a Hajek-style normalization improved both of the above approaches when covariate overlap between the target and the transported studies was problematic, and the stabilized nonparametric conditional density estimation approach performed especially well. The choice of weighting method is an active area of research in the general field of causal inference and will surely have implications for causally-interpretable meta-analysis as well.[46,49-51] The OM approach may be a more suitable approach in cases with poor overlap, as it allows for extrapolation of results. A deeper investigation of these methods and their respective strengths and weaknesses lies beyond this paper's scope but will be an important consideration in future work.

Using the causally-interpretable approach to characterize between-study heterogeneity is another important future direction. The advantage of considering heterogeneity through the causally-interpretable lens is that it allows investigators to decompose heterogeneity into two sources, covariate-shift heterogeneity (heterogeneity from differences between studies in the enrolled individuals) and design-related heterogeneity (such as protocols differing between studies). Although the approach as discussed in this paper only captures covariate-shift heterogeneity, it provides a principled foundation upon which one can add further sources of variation for a more realistic analysis while retaining a target population and causal interpretation. For example, variation between studies in treatment implementation is related to the consistency assumption in the above approach (and in much causal-inference work), while some other sources of variation do not relate to consistency and thus may be amenable to an approach like that of VanderWeele and Hernán.[38-40] Future work can focus on extending this approach to use more realistic assumptions and to use some combination of individual patient data and aggregate data.

HIGHLIGHTS

- Recent work has introduced a meta-analysis framework explicitly based in causal inference, which transports results from a meta-analysis to a specific target population. Although theoretically appealing, this work has been little applied in practice.
- This paper applies the recent meta-analysis framework to two case studies and compares the results with those from well-established meta-analysis methods. Strengths and weaknesses of the new approach are discussed.
- Causally-interpretable meta-analysis could radically enhance the field of meta-analysis as a whole, and understanding how to improve the methods in applications is an important step in this enhancement.

DATA AVAILABILITY STATEMENT

The data that support the findings of this study are potentially available on request. The data are not publicly available due to privacy or ethical restrictions.

## **References**

1. Egger M, Smith GD, Altman DG. *Systematic reviews in health care: meta-analysis in context*. 2nd ed: BMJ Publishing Group; 2001.

2. Schmid CH, Stijnen T, White I. *Handbook of Meta-analysis*. New York: Chapman & Hall/CRC; 2020.

3. Borenstein M, Hedges LV, Higgins JPT, Rothstein HR. *Introduction to Meta-Analysis*: Wiley; 2009.

4. Lin L, Chu H. Quantifying publication bias in meta-analysis. *Biometrics*. Sep 2018;74(3):785-794. doi:10.1111/biom.12817.

5. Murad MH, Montori VM, Ioannidis JPA, et al. How to read a systematic review and meta-analysis and apply the results to patient care: User's guides to the medical literature. *JAMA*. 2014;312(2):171-179. doi:10.1001/jama.2014.5559.

6. Schnitzer ME, Steele RJ, Bally M, Shrier I. A causal inference approach to network meta-analysis. *J Causal Inference*. 2016;4(2). doi:10.1515/jci-2016-0014.

7. Hasegawa T, Claggett B, Tian L, Solomon SD, Pfeffer MA, Wei L-J. The myth of making inferences for an overall treatment efficacy with data from multiple comparative studies via meta-analysis. *Stat Biosci*. 2017;9:284-297. doi:10.1007/s12561-016-9179-3.

8. Hernán MA, Robins JM. *Causal Inference: What If*. Chapman & Hall/CRC; 2020.

9. Dahabreh IJ, Petito LC, Robertson SE, Hernán MA, Steingrimsson JA. Toward causally interpretable meta-analysis: Transporting inferences from multiple randomized trials to a new target population. *Epidemiology*. 2020;31(3):334-344. doi:10.1097/EDE.0000000000001177.

10. Dahabreh IJ, Robertson SE, Petito LC, Hernán MA, Steingrimsson JA. Efficient and robust methods for causally interpretable meta-analysis: transporting inferences from multiple randomized trials to a target population. *arXiv preprint arXiv:1908.09230*. 2020.

11. Pearl J, Bareinboim E. External validity: from do-calculus to transportability across populations. *Stat Sci* 2014;29(4):579–595. doi:10.1214/14-STS486.

12. Colnet B, Mayer I, Chen G et al. Causal inference methods for combining randomized trials and observational studies: a review. *arXiv preprint arXiv:2011.08047*. 2020.

13. Degtiar I, Rose S. A review of generalizability and transportability. *arXiv preprint arXiv:2102.11904*. 2021.

14. Barker DH, Dahabreh IJ, Steingrimsson JA, et al. Causally Interpretable Meta-analysis: Application in Adolescent HIV Prevention. *Prevention Science* 2022;23(3):403–414. https://doi.org/10.1007/s11121-021-01270-3.

15. Debray TP, Moons KG, van Valkenhoef G, et al. Get real in individual participant data (IPD) meta-analysis: a review of the methodology. *Res Synth Methods* 2015;6(4):293-309. doi: 10.1002/jrsm.1160.

16. Riley RD, Lambert PC, Abo-Zaid G. Meta-analysis of individual participant data: rationale, conduct, and reporting. *BMJ* 2010;340:c221. doi: 10.1136/bmj.c221.

17. Stewart LA, Clarke M, Rovers M, et al. Preferred Reporting Items for Systematic Review and Meta-Analyses of individual participant data: the PRISMA-IPD Statement. *JAMA* 2015;313(16):1657-65. doi: 10.1001/jama.2015.3656.

18. Zhou J, Hodges JS, Suri MFK, et al. A Bayesian hierarchical model estimating CACE in meta-analysis of randomized clinical trials with noncompliance. *Biometrics* 2019;75(3):978-87. doi: 10.1111/biom.13028.

19. Zhou J, Hodges JS, Chu H. Rejoinder to "CACE and meta-analysis (letter to the editor)" by Stuart Baker. *Biometrics* 2020;76(4):1385-89. doi: 10.1111/biom.13239.

20. Zhou J, Hodges JS, Chu H. A Bayesian Hierarchical CACE Model Accounting for Incomplete Noncompliance With Application to a Meta-analysis of Epidural Analgesia on Cesarean Section. *Journal of the American Statistical Association* 2021:1-13. doi: 10.1080/01621459.2021.1900859.

21. Zhou T, Zhou J, Hodges JS, et al. Estimating the Complier Average Causal Effect in a Meta-analysis of Randomized Clinical Trials with Binary Outcomes Accounting for Noncompliance: A Generalized Linear Latent and Mixed Model Approach. *Am J Epidemiol* 2021 doi: 10.1093/aje/kwab238.

22. Domínguez Islas C, Rice KM. Addressing the estimation of standard errors in fixed effects meta-analysis. *Stat Med*. 2018;37(11):1788-1809. doi:10.1002/sim.7625.

23. Rice K, Higgins JPT, Lumley T. A re-evaluation of fixed effect(s) meta-analysis. *J. R. Stat. Soc. Ser. A*. 2018;181(1):205-227. doi:10.1111/rssa.12275.

24. Higgins JP, Thompson SG. Quantifying heterogeneity in a meta-analysis. *Stat Med*. 2002;21(11):1539-1558. doi: 10.1002/sim.1186.

25. Higgins JP, Thompson SG, Deeks JJ, Altman DG. Measuring inconsistency in meta-analyses. *BMJ*. 2003;327(7414):557-560. doi: 10.1136/bmj.327.7414.557.

26. Lin L, Chu H, Hodges JS. Alternative measures of between-study heterogeneity in meta-analysis: Reducing the impact of outlying studies. *Biometrics*. 2017;73(1):156-166. doi: 10.1111/biom.12543.

27. Higgins JPT, Thompson SG, Spiegelhalter DJ. A re-evaluation of random-effects meta-analysis. *J. R. Stat. Soc. Ser. A*. 2009;172(1):137-159. doi:10.1111/j.1467-985X.2008.00552.x.

28. Huang Y, Mao C, Yuan J, et al. Distribution and epidemiological characteristics of published individual patient data meta-analyses. *PLoS One*. 2014;9(6):e100151. doi:10.1371/journal.pone.0100151.

29. Tipton E, Pustejovsky JE, Ahmadi H. A history of meta-regression: technical, conceptual, and practical developments between 1974 and 2018. *Research Synthesis Methods*. 2018;10(2):161-179. doi:10.1002/jrsm.1338.

30. Taylor JA, Kowalski SM, Polanin JR, et al. Investigating science education effect sizes: implications for power analyses and programmatic decisions. *AERA Open*. 2018;4(3):1-10. doi: 10.1177/2332858418791991.

31. Williams R, Citkowicz M, Miller DI, et al. Heterogeneity in mathematics intervention effects: evidence from a meta-analysis of 191 randomized experiments. *Journal of Research on Educational Effectiveness*. 2022;15(3):584-634. doi:10.1080/19345747.2021.2009072.

32. van Lissa C. Small sample meta-analyses: exploring heterogeneity using MetaForest. In: van de Schoot R, Miočević M, ed. *Small Samplie Size Solutions: A Guide for Applied Researchers and Practitioners*. 1st ed. London: Routledge; 2020:186-202.

33. Cinar O, Umbanhowar J, Hoeksema JD, et al. Using information-theoretic approaches for model selection in meta-analysis. *Research Synthesis Methods*. 2021;12(4):537-556. doi: 10.1002/jrsm.1489.

34. Tanner-Smith EE and Tipton E. Robust variance estimation with dependent effect sizes: practical considerations including a software tutorial in Stata and SPSS. *Research Synthesis Methods*. 2014;5(1):13-30. doi:10.1002/jrsm.1091.

35. Kabali C, Ghazipura M. Transportabilty in network meta-analysis. *Epidemiology*. 2016;27(4):556-561. doi:10.1097/EDE.0000000000000475.

36. Sobel M, Madigan D, Wang W. Causal inference for meta-analysis and multi-level data structures, with application to randomized studies of Vioxx. *Psychometrika*. 2017;82(2):459-474. doi:10.1007/s11336-016-9507-z.

37. Dahabreh IJ, Robins JM, Haneuse SJ-PA. Sensitivity analysis using bias functions for studies extending inferences from a randomized trial to a target population. *arXiv preprint arXiv:1905.10684*. 2019.

38. VanderWeele TJ, Hernán MA. Causal inference under multiple versions of treatment. *J Causal Inference*. 2013;1(1):1-20. doi:10.1515/jci-2012-0002.

39. Hasegawa RB, Deshpande SK, Small DS, Rosenbaum PR. Causal inference with two versions of treatment. *Journal of Educational and Behavioral Statistics*. 2020;45(4):426-445. doi:10.3102/1076998620914003.

40. Heiler P, Knaus MC. Effect or treatment heterogeneity? Policy evaluation with aggregated and disaggregated treatments. *arXiv preprint arXiv:2110.01427*. 2021.

41. Ishak KJ, Proskorovsky I, Benedict A. Simulation and Matching-Based Approaches for Indirect Comparison of Treatments. *Pharmacoeconomics*. 2015;33:537–549. doi:10.1007/s40273-015-0271-1.

42. Phillippo DM, Dias S, Ades AE, et al. Multilevel network meta-regression for population-adjusted treatment comparisons, *J. R. Stat. Soc. Ser. A*. 2020;183(3):1189-1210. doi: 10.1111/rssa.12579.

43. Signorovitch JE, Wu EQ, Yu AP, et al. Comparative Effectiveness Without Head-to-Head Trials: A Method for Matching-Adjusted Indirect Comparisons Applied to Psoriasis Treatment with Adalimumab or Etanercept. *Pharmacoeconomics*. 2010;28(10):935-945. doi: 10.2165/11538370-000000000-00000.

44. Saul BC, Hudgens MG. The calculus of M-estimation in R with geex. *J Stat Softw*. 2020;92(2):1-15. doi:10.18637/jss.v092.i02.

45. Hayfield T, Racine JS. Nonparametric econometrics: the np package. *J Stat Softw*. 2008;27(5). doi:10.18637/jss.v027.i05.

46. Chattopadhyay A, Hase CH, Zubizarreta JR. Balancing vs modeling approaches to weighting in practice. *Stat Med*. 2020;39(24):3227-3254. doi:10.1002/sim.8659.

47. Benkhadra K, Alahdab F, Tamhane S, et al. Real-time continuous glucose monitoring in type 1 diabetes: a systematic review and individual patient data meta-analysis. *Clin Endocrinol (Oxf)*. 2017;86(3):354-360. doi:10.1111/cen.13290.

48. Balduzzi S, Rücker G, Schwarzer G. How to perform a meta-analysis with R: a practical tutorial. *Evid Based Ment Health*. 2019;22(4):153-160. doi:10.1136/ebmental-2019-300117.

49. Kranker K, Blue L, Forrow LV. Improving effect estimates by limiting the variability in inverse propensity score weights. *Am Stat*. 2021;75(3): 276-287. doi: 10.1080/00031305.2020.1737229.

50. Crump RK, Hotz VJ, Imbens GW, Mitnik OA. Dealing with limited overlap in estimation of average treatment effects. *Biometrika.* 2009;96(1):187-199. doi: 10.1093/biomet/asn055.

51. Chan KCG, Yam SCP, Zhang Z. Globally efficient non-parametric inference of average treatment effects by empirical balancing calibration weighting. *J R Stat Soc Series B Stat Methodol*. 2016; 78(3):673-700. doi: 10.1111/rssb.12129.

**Table 1. Applying Estimators to Real-time Continuous Glucose Monitoring Data.**

| Target Study | Observed TE | OM | IPW | IPW-h | np IPW | np IPW-h | FE MA | RE MA | Meta-reg |
|---|---|---|---|---|---|---|---|---|---|
| 2 | -0.619 | -0.293 | -0.864 | -0.347 | -0.245 | -0.192 | -0.119 | -0.146 | -0.504 |
| 3 | -0.388 | -0.367 | -1.582 | -0.872 | 0.333 | -0.314 | -0.150 | -0.191 | -0.631 |
| 4 | -0.614 | -0.219 | -10.241 | -1.003 | 0.422 | -0.210 | -0.139 | -0.166 | -0.023 |
| 5 | -0.254 | -0.072 | 1.600 | -0.434 | -0.042 | -0.208 | -0.144 | -0.206 | -0.104 |
| 7 | 0.228 | -0.345 | -0.388 | -0.336 | 0.068 | -0.348 | -0.208 | -0.269 | -0.378 |
| 9 | 0.174 | -0.139 | -4.087 | -0.395 | -1.385 | -0.161 | -0.208 | -0.265 | -0.437 |
| 10 | -0.130 | -0.157 | -0.554 | -0.099 | -0.492 | -0.228 | -0.177 | -0.231 | 0.302 |
| 12 | -0.303 | -0.248 | -0.327 | -0.254 | 0.014 | -0.153 | -0.147 | -0.198 | -0.287 |
| Avg Abs Diff | | 0.236 | 2.281 | 0.317 | 0.593 | 0.264 | 0.293 | 0.288 | 0.345 |
| St Abs Diff | | 1.394 | 13.407 | 1.826 | 3.579 | 1.595 | 1.718 | 1.728 | 2.242 |

Abbreviations: Observed TE = Observed treatment effect in the target study. OM = Outcome model. IPW = Inverse probability weighting. np IPW = Inverse probability weighting using nonparametric conditional density estimation. -h suffix indicates Hajek normalization. FE MA = Fixed-effect meta-analysis. RE MA = Random-effects meta-analysis. Meta-reg = aggregate meta-regression.

**Table 2. Applying Estimators to Chiropractic Data.**

| Target Study | Observed TE | OM | IPW | IPW-h | np IPW | np IPW-h | FE MA | RE MA | Meta-reg |
|---|---|---|---|---|---|---|---|---|---|
| 1 | -0.241 | -0.359 | -0.304 | -0.281 | -0.267 | -0.262 | -0.214 | -0.219 | -0.360 |
| 3 | 0.358 | -0.375 | -0.482 | -0.475 | -0.451 | -0.458 | -0.342 | -0.342 | -0.456 |
| 4 | -0.301 | -0.528 | -0.324 | -0.410 | -0.372 | -0.434 | -0.201 | -0.205 | 0.461 |
| 5 | -0.408 | -0.102 | -0.127 | -0.125 | -0.231 | -0.164 | -0.192 | -0.193 | -0.008 |
| 6 | -0.504 | 0.014 | -0.005 | -0.014 | 0.072 | 0.024 | -0.178 | -0.178 | 0.037 |
| 7 | -0.341 | -0.248 | -0.230 | -0.222 | -0.248 | -0.236 | -0.195 | -0.198 | -0.278 |
| Avg Abs Diff | | 0.333 | 0.303 | 0.312 | 0.292 | 0.308 | 0.253 | 0.250 | 0.450 |
| St Abs Diff | | 2.160 | 1.973 | 2.033 | 1.902 | 2.003 | 1.648 | 1.632 | 2.924 |

Abbreviations: Observed TE = Observed treatment effect in the target study. OM = Outcome model. IPW = Inverse probability weighting. np IPW = Inverse probability weighting using nonparametric conditional density estimation. -h suffix indicates Hajek normalization. FE MA = Fixed-effect meta-analysis. RE MA = Random-effects meta-analysis. Meta-reg = aggregate meta-regression.

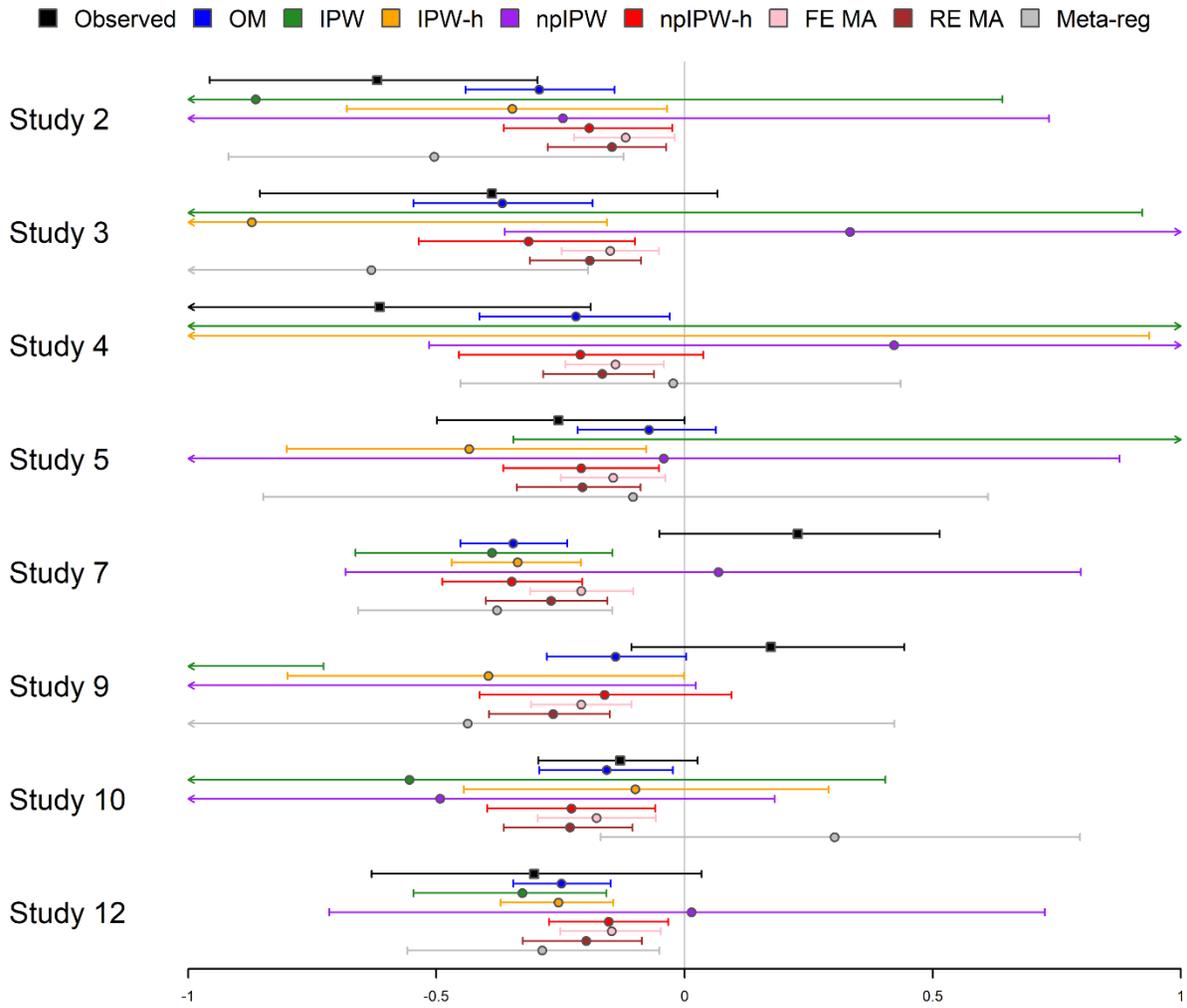

Figure 1. Estimated Treatment Effects with 95% CI, Glucose Monitoring Data.

Figure 2. Estimated Treatment Effects with 95% CI, Chiropractic Data.